\documentclass[twocolumn,showpacs,preprintnumbers,amsmath,amssymb,prl]{revtex4}
\usepackage{graphicx}
\usepackage{dcolumn}
\usepackage{epsf}

\begin{document}
\title{Limits of the upper critical field in dirty two-gap superconductors}
\author{A. Gurevich}
\affiliation{National High Magnetic Field Laboratory, Florida State
University, Tallahassee, FL 32310, USA}
\date{\today}

\begin{abstract}
An overview of the theory of the upper critical field in dirty
two-gap superconductors, with a particular emphasis on MgB$_2$ is
given. We focus here on the maximum $H_{c2}$ which may be achieved
by increasing intraband scattering, and on the limitations imposed
by weak interband scattering and paramagnetic effects. In
particular, we discuss recent experiments which have recently
demonstrated ten-fold increase of $H_{c2}$ in dirty carbon-doped
films as compared to single crystals, so that the $H_{c2}(0)$
parallel to the ab planes may approach the BCS paramagnetic limit,
$H_p[T] = 1.84T_c[K] \simeq 60-70T$. New effects produced by weak
interband scattering in the two-gap Ginzburg-Landau equations and
$H_{c2}(T)$ in ultrathin MgB$_2$ films are addressed.

\end{abstract}
\maketitle

\section{Introduction}

It is now well established that superconductivity in MgB$_2$ with
the unexpectedly high critical temperature $T_c\approx 40$K
\cite{mgb2}, is due to strong electron-phonon interaction with
in-plane boron vibration modes. Extensive ab-initio calculations
\cite{tg1,tg2,tg3}, along with many experimental evidences from STM,
point contact, and Raman spectroscopy, heat capacity, magnetization
and rf measurements \cite{rev1,rev2} unambiguously indicate that
MgB$_2$ exhibits two-gap s-wave superconductivity
\cite{suhl,moskal}. MgB$_2$ has two distinct superconducting gaps:
the main gap $\Delta_\sigma (0)\approx 7.2$mV, which resides on the
2D cylindrical parts of the Fermi surface formed by in-plane
$\sigma$ antibonding $p_{xy}$ orbitals of B, and the smaller gap
$\Delta_\pi (0)\approx 2.3$mV on the 3D tubular part of the Fermi
surface formed by out-of-plane $\pi$ bonding and antibonding $p_z$
orbitals of B.

The discovery of MgB$_2$ has renewed interest in new effects of
two-gap superconductivity, motivating different groups to take
closer looks at other known materials, such as YNi$_2$B$_2$C and
LuNi$_2$B$_2$C borocarbides \cite{borocarb} Nb$_3$Sn \cite{nb3sn},
or NbSe$_2$ \cite{nb2se}, heavy-fermion \cite{heavyferm} and organic
\cite{organsc} superconductors, for which evidences of the two gap
behavior have been reported. However, several features of MgB$_2$
set it apart from other two-gap superconductors. Not only does
MgB$_2$ have the highest $T_c$ among all non-cuprate
superconductors, it also has two coexisting order parameters
$\Psi_\sigma=\Delta_\sigma\exp(i\theta_1)$ and
$\Psi_\pi=\Delta_\pi\exp(i\theta_2)$, which are {\it weakly
coupled}. The latter is due to the fact that the $\sigma$ and $\pi$
bands are formed by two orthogonal sets of in-plane and out-of-plane
atomic orbitals of boron, so all overlap integrals, which determine
matrix elements of interband coupling and interband impurity
scattering are strongly reduced \cite{mazimp}. This feature can
result in new effects, which are very important both for the physics
and applications of MgB$_2$. Indeed, two weakly coupled gaps result
in {\it intrinsic Josephson effect}, which can manifest itself in
low-energy interband Josephson plasmons (the Legget mode)
\cite{legget} with frequencies smaller than $\Delta_\pi/\hbar$.
Moreover, strong static electric fields and currents can decouple
the bands due to formation of interband textures of $2\pi$ planar
phase slips in the phase difference $\theta (x)=\theta_1-\theta_2$
\cite{text1,text2} well below the global depairing current. In turn,
the weakness of interband impurity scattering makes it possible to
radically increase the upper critical field $H_{c2}$ by selective
alloying of Mg and B sites with nonmagnetic impurities.

Despite the comparatively high $T_c$, the upper critical field of
MgB$_2$ single crystals is rather low and anisotropic with
$H_{c2}^{\perp}(0)\simeq 3-5T$ and $H_{c2}^{||}(0)\simeq 15-19T$ of
\cite{rev1,rev2}, where the indices $\perp$ and $||$ correspond to
the magnetic field {\bf H} perpendicular and parallel to the ab
plane, respectively. Since these $H_{c2}$ values are significantly
lower than $H_{c2}(0)\simeq 30$T for Nb$_3$Sn \cite{orlando,arno},
there had been initial scepticism about using MgB$_2$ as a
high-field superconductor, until several groups undertook the
well-established  procedure of $H_{c2}$ enhancement by alloying
MgB$_2$ with nonmagnetic impurities. The results of high-field
measurements on dirty MgB$_2$ films and bulk samples has shown up to
ten-fold increase of $H_{c2}^{\perp}$ as compared to single crystals
\cite{h1,h2,h3,h4,h5,h6,h7,h8,h9,h10,h11,h12}, particularly in
carbon-doped thin films \cite{h9} made by hybrid physico-chemical
vapor deposition \cite{penn}. This unexpectedly strong enhancement
of $H_{c2}(T)$ results from its anomalous upward curvature, rather
different from that of $H_{c2}(T)$ for one-gap dirty superconductors
\cite{agork,dege,whh,maki}. As shown in Fig. 1, $H_{c2}$ of MgB$_2$
C-doped films has already surpassed $H_{c2}$ of Nb$_3$Sn, which
could make cheap and ductile MgB$_2$ an attractive material for high
field applications \cite{ap1}.

\begin{figure}          
\epsfxsize= 0.8\hsize \centerline{ \vbox{ \epsffile{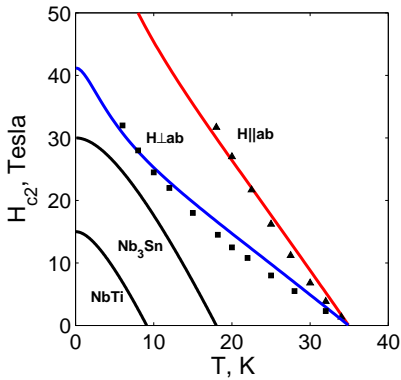} }}
\vskip \baselineskip \caption{$H_{c2}(T)$ for carbon-doped MgB$_2$
films \cite{h9} in comparison with NbTi and Nb$_3$Sn. The red and
blue lines show fits from Eq. (\ref{final}) with $g=0.045$}.
\label{fig.1}
\end{figure}

This radical enhancement of $H_{c2}$ shown in Fig. 1 is indeed
assisted by the features of two-gap superconductivity in MgB$_2$.
Fig. 2 gives another example of $H_{c2}(T)$ for a fiber-textured
film \cite{h6}, which exhibits an upward curvature of $H_{c2}(T)$
for $H||c$. This behavior of $H_{c2}(T)$ and the anomalous
temperature-dependent anisotropy ratio
$\Gamma(T)=H_{c2}^{||}(T)/H_{c2}^{\perp}(T)$ are different from that
of the one-gap theory in which the $H_{c2}(T)$ has a downward
curvature, while the slope $H_{c2}'=dH_{c2}/dT$ at $T_c$ is
proportional to the normal state residual resistivity $\rho_n$, and
$H_{c2}(0)=0.69T_cH_{c2}'$ \cite{agork,dege,whh,maki}. However, the
behavior of $H_{c2}(T)$ in MgB$_2$ can be explained by the two-gap
theory in the dirty limit based on either Usadel equations
\cite{ag,gk} or Eliashberg equations \cite{borocarb,carb}.

\begin{figure}          
\epsfxsize= 0.85\hsize \centerline{ \vbox{ \epsffile{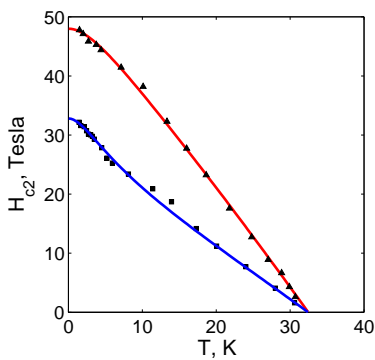} }}
\vskip \baselineskip \caption{$H_{c2}(T)$ of a fiber-textured
MgB$_2$ film \cite{h6} both parallel (triangles) and parallel
(squares) to the ab planes. The solid lines show calculations from
Eq. (\ref{final}) with $g=0.065$, $D_\pi\ll D_\sigma^{(ab)}$ for
$H||c$ and $D_\pi=0.19(D_\sigma^{(c)}D_\sigma^{(ab)})^{1/2}$ for
$H\perp c$.} \label{fig.2}
\end{figure}
\begin{figure}          
\epsfxsize= 0.85\hsize \centerline{ \vbox{ \epsffile{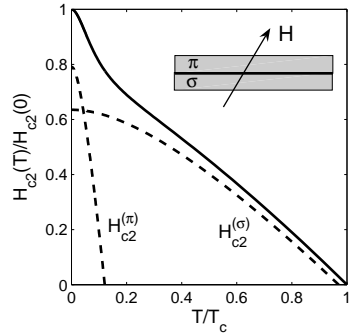} }}
\vskip \baselineskip \caption{The mechanism of the upward curvature
of $H_{c2}(T)$ illustrated by the bilayer toy model shown in the
inset. The dashed curves show $H_{c2}(T)$ calculated for $\sigma$
and $\pi$ films in the one-gap dirty limit with the BCS coupling
constants $\lambda_\sigma=0.81$, $\lambda_\pi=0.285$, and $D_\pi =
0.1D_\sigma$. The solid curve shows $H_{c2}(T)$ calculated from Eq.
(\ref{hc2}) of the two-gap dirty limit theory for the BCS matrix
constants from Ref. \cite{goluba}} \label{fig.3}
\end{figure}

The behavior of $H_{c2}(T)$ can be qualitatively understood using a
simple bilayer model shown in Fig. 3, which captures the physics of
two-gap superconductivity in MgB$_2$, and suggests ways by which
$H_{c2}$ can be further increased. Indeed, MgB$_2$ can be mapped
onto a bilayer in which two thin films corresponding to $\sigma$ and
$\pi$ bands are separated by a Josephson contact, which models the
interband coupling. The global $H_{c2}(T)$ of the such
weakly-coupled bilayer is mostly determined by the film with the
highest $H_{c2}$, even if $T_c^{(\sigma)}$ and $T_c^{(\pi)}$ are
very different. For example, if the $\pi$ film is much dirtier than
the $\sigma$ film then $H_{c2}^{(\sigma)}$ dominates at higher T,
but at lower temperatures the $\pi$ film takes over, resulting in
the upward curvature of $H_{c2}(T)$. If the $\sigma$ film is
dirtier, the $\pi$ film only results in a slight shift of the
$H_{c2}$ curve and a reduction of the slope $H_{c2}^\prime$ near
$T_c$.

The bilayer model also clarifies the anomalous angular dependence of
$H_{c2}(\alpha,T)$ for ${\bf H}$ inclined by the angle $\alpha$ with
respect to the c-axis (parallel to the film normal in Fig. 3)
\cite{note}. In this case both $H_{c2}^{(\sigma)}(\alpha,T)$ and
$H_{c2}^{(\pi)}(\alpha,T)$ depend on $\alpha$ according to the
temperature-independent one-gap scaling
$H_{c2}(\alpha)=H_{c2}(0,T)/\sqrt{\cos^2\alpha +
\epsilon\sin^2\alpha}$ \cite{gork,bul}, but with very different
effective mass ratios $\epsilon=m_{ab}/m_c$ for each film. Because
the $\sigma$ band is much more anisotropic than the $\pi$ band,
$\epsilon_\sigma\ll 1$, and $\epsilon_\pi\sim 1$ \cite{anz1,anz2},
the one-gap angular scaling for the global $H_{c2}(\alpha,T)$ breaks
down. For example, in the case shown in Fig. 3, $H_{c2}(T)$ is
anisotropic at higher T, but at lower T, the nearly isotropic $\pi$
band reduces the overall anisotropy of $H_{c2}$, so the ratio
$\Gamma(T)=H_{c2}^{||}(T)/H_{c2}^{\perp}(T)$ {\it decreases} as T
decreases. This is characteristic of many dirty MgB$_2$ films like
the one shown in Fig. 2, for which the $\pi$ band is typically much
dirtier than the $\sigma$ band. By contrast, in clean MgB$_2$ single
crystals $\Gamma(T)$ increases from $\simeq 2-3$ near $T_c$ to
$\simeq 5-6$ at $T\ll T_c$
\cite{a1,a2,a3,a4,a5,a6,a7,a8,a9,a10,a11}. This behavior was
explained by two-gap effects in the clean limit \cite{vgk,dahm}.

Fig. 3 suggests that $H_{c2}(T)$ of MgB$_2$ can be significantly
increased at low T by making the $\pi$ band much dirtier than the
main $\sigma$ band. This could be done by disordering the Mg
sublattice, thus disrupting the $p_z$ boron out-of-plane orbitals,
which form the $\pi$ band. Achieving high $H_{c2}$ requires that
both $\sigma$ and $\pi$ bands are in the dirty limit. Yet, making
the $\pi$ band much dirtier than the $\sigma$ band provides a "free
boost" in $H_{c2}$ without too much penalty in $T_c$ suppression due
to pairbreaking interband scattering or band depletion due to doping
\cite{depl1,depl2}. In fact, the interband scattering is weak for
the same reason that $\Psi_\sigma$ and $\Psi_\pi$ are weakly
coupled, which may enable alloying MgB$_2$ with more impurities to
achieve higher $H_{c2}$. Systematic incorporation of impurities in
MgB$_2$ has not been yet achieved because the complex substitutional
chemistry of MgB$_2$ is still poorly understood
\cite{chem1,chem2,chem3,chem4}. Several groups have reported a
significant increase in $H_{c2}$ by irradiation with protons
\cite{prot}, neutrons \cite{neut1,neut2} or heavy ions \cite{asu},
but so far the carbon impurities have been the most effective to
provide the huge $H_{c2}$ enhancement shown in Figs. 1 and 2. The
effect of carbon on different superconducting properties can be
rather complex \cite{carb1,carb2,carb3} and still far from being
fully understood. Yet given the indisputable benefits of carbon
alloying, one can pose the basic question: how far can $H_{c2}$ be
further increased?

The bilayer model suggests that $H_{c2}$ increases if intraband
scattering is enhanced. However, because intraband impurity
scattering causes an admixture of pairbreaking interband scattering,
the first question is to what extent weak interband scattering in
MgB$_2$ can limit $H_{c2}$. Another important question is how far is
the observed $H_{c2}$ from the paramagnetic limit $H_p$. In the BCS
theory  $H_p$ is defined by the condition: $\mu_BH_p^{BCS} =
\Delta/\sqrt{2}$, or $H_p^{BCS}[T] = 1.86T_c[K]$ \cite{par3}, where
$\mu_B$ is the Bohr magneton. For $T_c=35K$, this yields $H_p=65$T,
not that far from the zero-field $H_{c2}^{||}(0)$ in Figs. 1 and 2.
However, the BCS model underestimates $H_p$, which is significantly
enhanced by strong electron-phonon coupling \cite{par4}:
    \begin{equation}
    H_p\simeq (1+\lambda_{ep})H_p^{BCS},
    \label{paraeli}
    \end{equation}
where $\lambda_{ep}$ is the electron-phonon constant. Taking
$\lambda_{ep}\approx 1$ for the $\sigma$ band \cite{tg1,tg2}, we
obtain $H_p\sim 130$T, so there still a large room for increasing
$H_{c2}$ by optimizing the intra and interband impurity scattering.
For instance, increasing $H_{c2}'$ to a rather common for many high
field superconductors value of 2T/K  (much lower than $H_{c2}'\simeq
5-14$T/K for PbMo$_6$S$_8$ \cite{pb}) could drive $H_{c2}$ of
MgB$_2$ with $T_c\simeq 35K$ above 70T. In the following we give a
brief overview of recent results in the theory of dirty two-gap
superconductors focusing on new effects brought by weak interband
scattering and paramagnetic effects. The main conclusion is that,
although interband scattering in MgB$_2$ is indeed weak, it cannot
be neglected in calculations of $H_{c2}(T)$. We will also address
the crossover from the orbitally-limited to the paramagnetically
limited $H_{c2}$ in a two-gap superconductor.

\section{Tho-gap superconductors in the dirty limit}

We regard MgB$_2$ as a dirty anisotropic superconductor with two
sheets 1 and 2 of the Fermi surface on which the superconducting
gaps take the values $\Delta_1$ and $\Delta_2$, respectively
(indices 1 and 2 correspond to $\sigma$ and $\pi$ bands). Although
the $\sigma$ band is anisotropic, MgB$_2$ is not a layered material
\cite{layer1,layer2}, so the continuum BCS theory is applicable
because the c-axis coherence length $\xi_c$ is much longer than the
spacing between the boron planes $\sim 3.5\AA$. Indeed, even for
$H_{c2}^{\perp}(0)=40$T and $H_{c2}^{||}(0)=60$T in Fig. 1, the
anisotropic Ginzburg-Landau (GL) theory \cite{gork} gives
$\xi_c=(\phi_0H_{c2}^\perp/2\pi)^{1/2}/H_{c2}^{||}\approx 19\AA$.
Strong coupling in MgB$_2$ should be described by the Eliashberg
equations \cite{carb}, but we consider here manifestations of intra
and interband scattering and paramagnetic effects in $H_{c2}$ using
the more transparent two-gap Usadel equations \cite{ag}
    \begin{eqnarray}
    \omega f_1-\frac{D_1^{\alpha\beta}}{2}[g_1\Pi_\alpha\Pi_\beta f_1-f_1\nabla_\alpha\nabla_\beta g_1]
    \nonumber \\
    =\Psi_1g_1+\gamma_{12}(g_1 f_2-g_2 f_1)\qquad
    \label{uz1} \\
    \omega f_2-\frac{D_2^{\alpha\beta}}{2}[g_2\Pi_\alpha\Pi_\beta f_2-f_2\nabla_\alpha\nabla_\beta g_2]
    \nonumber \\
    =\Psi_2g_2+\gamma_{21}(g_2 f_1 -g_1 f_2),\qquad
    \label{uz2}
    \end{eqnarray}
Here the Usadel Green's functions $f_m({\bf r},\omega)$ and
$g_m({\bf r},\omega)$ in the m-th band depend on ${\bf r}$ and the
Matsubara frequency $\omega = \pi T(2n+1)$, $D_m^{\alpha\beta}$ are
the intraband diffusivities due to nonmagnetic impurity scattering,
$2\gamma_{mm'}$ are the interband scattering rates, ${\bf
\Pi}=\nabla + 2\pi i {\bf A}/\phi_0$, ${\bf A}$ is the vector
potential, and $\phi_0$ is the flux quantum. Eqs. (\ref{uz1}) and
(\ref{uz2}) are supplemented by the equations for the order
parameters $\Psi_m=\Delta_m\exp(i\varphi_m)$,
    \begin{equation}
    \Psi_m=2\pi T\sum_{\omega > 0}^{\omega_D}\sum_m\lambda_{mm'}f_{m'}({\bf r},
    \omega),
    \label{d}
    \end{equation}
normalization condition $|f_m |^2+g_m^2=1$, and the supercurrent
density
    \begin{equation}
    J^\alpha=-2\pi eT Im\sum_\omega\sum_mN_mD_m^{\alpha\beta}f_m^*\Pi_\beta
    f_m.
    \label{j}
    \end{equation}
Here $N_m$ is the partial electron density of states for both spins
in the m-th band, and $\alpha$ and $\beta$ label Cartesian indices.
Eqs. (\ref{d}) contains the matrix of the BCS coupling constants
$\lambda_{mm'}=\lambda_{mm'}^{(ep)}-\mu_{mm'}$, where
$\lambda_{mm'}^{(ep)}$ are electron-phonon constants, and
$\mu_{mm'}$ is the Coulomb pseudopotential. The diagonal terms
$\lambda_{11}$ and $\lambda_{22}$ quantify intraband pairing, and
$\lambda_{12}$ and $\lambda_{21}$ describe interband coupling.
Hereafter, the following {\it ab initio} values
$\lambda_{\sigma\sigma}\approx 0.81$, $\lambda_{\pi\pi}\approx
0.285$, $\lambda_{\sigma\pi}\approx 0.119$, and
$\lambda_{\pi\sigma}\approx 0.09$ \cite{goluba} are used. There are
also the symmetry relations:
    \begin{equation}
    N_1\lambda_{12}=N_2\lambda_{21},\qquad N_1\gamma_{12}=N_2\gamma_{21}
    \label{inter}
    \end{equation}
where $N_\pi\approx 1.3N_\sigma$ for MgB$_2$. Solutions of Eqs.
(\ref{uz1})-(\ref{inter}) minimize the following free energy $\int
Fd^3{\bf r}$ \cite{text2}:
    \begin{equation}
    F=\frac{1}{2}\sum_{mm^\prime}N_m\Psi_m\Psi_m^*\lambda_{mm^\prime}^{-1}+F_1+F_2+F_i
    \label{FF}
    \end{equation}
Here $F_1$ and $F_2$ are intraband contributions,
    \begin{eqnarray}
    F_m=2\pi T\sum_{\omega > 0}N_m[(\omega (1-g_m)- \\ \nonumber
    Re(f_m^*\Delta_m) +D_m^{\alpha\beta}[\Pi_\alpha f_m\Pi_\beta^* f_m^*+\nabla_\alpha g_m\nabla_\beta
    g_m]/4
    \label{Fm}
    \end{eqnarray}
and $F_i$ is due to interband scattering \cite{agg}:
    \begin{equation}
    F_i=2\pi q T\sum_{\omega > 0} [1-g_1 g_2 -\mbox{Re}(f_1^*f_2)],
    \label{Fi}
    \end{equation}
where $2q= N_1\gamma_{12}+N_2\gamma_{21}$. The Usadel equations
result from $\delta F/\delta f_m^*=0$, $\partial F/\partial
\Psi_m^*=0$, and ${\bf J}=-c\delta F/\delta {\bf A}$. Taking $f_m =
\sin \alpha_m$ and $g_m=\cos\alpha_m$, we obtain
    \begin{eqnarray}
    \omega\sin\alpha_1+\gamma_{12}\sin(\alpha_1-\alpha_2)=\Delta_1\cos\alpha_1,
    \label{thet1} \\
    \omega\sin\alpha_2+\gamma_{21}\sin(\alpha_2-\alpha_1)=\Delta_2\cos\alpha_2.
    \label{thet2}
    \end{eqnarray}
These coupled equations along with Eq. (\ref{d}) define the two-gap
uniform states for $J=0$.

\section{Critical temperature}

Eqs. (\ref{uz1}) and (\ref{uz2}) give the well-known results for
$T_c$ in two-gap superconductors \cite{suhl,moskal,imp1,imp2}. For
negligible interband scattering, substitution of
$f_1=\Delta_1/\omega$ and $f_2=\Delta_2/\omega$ into Eq. (\ref{d})
yields:
    \begin{equation}
    T_{c0}=1.14\hbar\omega_D\exp[-(\lambda_+-\lambda_0)/2w],
    \label{suhl}
    \end{equation}
where $\lambda_{\pm}=\lambda_{11}\pm\lambda_{22}$,
$w=\lambda_{11}\lambda_{22}-\lambda_{12}\lambda_{21}$, and
$\lambda_0=(\lambda_-^2+4\lambda_{12}\lambda_{21})^{1/2}$. The
interband coupling increases $T_{c0}$ as compared to noninteracting
bands ($\lambda_{12}=\lambda_{21}=0$), while intraband impurity
scattering does not affect $T_{c0}$, in accordance with the Anderson
theorem. Solving the linearized Eqs. (\ref{uz1}) and (\ref{uz2})
with $\gamma_{mm'}\neq 0$, gives $T_c$ with the account of
pairbreaking interband scattering:
    \begin{eqnarray}
    U\left(\frac{g}{t_c}\right)=-\frac{(\lambda_0+w\ln t_c)\ln t_c}{p+w\ln t_c},\qquad
    \label{tc} \\
    2p=\lambda_0+[\gamma_-\lambda_- -2\lambda_{21}\gamma_{12}-2\lambda_{12}\gamma_{21}]/\gamma_+,
    \label{p}\\
    U(x)=\psi(1/2+x)-\psi(1/2),\qquad\qquad
    \end{eqnarray}
where $t_c=T_c/T_{c0}$ and $\gamma_\pm=\gamma_{12}\pm\gamma_{21}$,
$g=\gamma_+/2\pi T_{c0}$, and $\psi(x)$ is a digamma function. The
dependence of $T_c$ on the interband scattering parameter g is shown
in Fig. 4. As $g\to\infty$, Eqs. (\ref{tc}) and (\ref{p}) give
$T_c\to T_{c0}\exp (-p/w)$, and for $g\ll 1$, we have
    \begin{equation}
    T_c=T_{c0}-
    \frac{\pi}{8\lambda_0}[\lambda_0\gamma_++\lambda_-\gamma_--2\lambda_{21}\gamma_{12}-2\lambda_{12}\gamma_{21}]
    \label{tcsmall}
    \end{equation}
This formula can be used to extract the interband scattering rates
from the small shift of $T_c$ \cite{maria}. However, as shown below,
even weak interband scattering can significantly change the behavior
of $H_{c2}(T)$, so it cannot be neglected even though $g\ll 1$.

\begin{figure}          
\epsfxsize= 0.8\hsize \centerline{ \vbox{ \epsffile{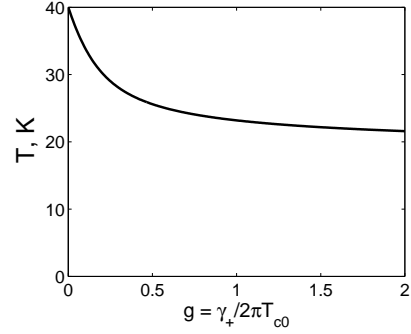} }}
\vskip \baselineskip \caption{Dependence of the critical temperature
$T_c$ on the interband scattering parameter g calculated from Eq.
(\ref{tc}) with the BCS matrix constants $\lambda_{mn}$ from Ref.
\cite{goluba} } \label{fig.4}
\end{figure}

\section{Upper critical field for ${\bf H}\| c$}

$H_{c2}$ along the c-axis is the maximum eigenvalue of the
linearized Eqs. (\ref{uz1}) and (\ref{uz2}):
    \begin{eqnarray}
    (\omega \pm i\mu_BH)f_1-\frac{D_1}{2}\Pi^2f_1=\Delta_1+(f_2-f_1)\gamma_{12},
    \label{uzl1} \\
    (\omega \pm i\mu_BH)f_2-\frac{D_2}{2}\Pi^2f_2=\Delta_2+(f_1-f_2)\gamma_{21},
    \label{uzl2}
    \end{eqnarray}
Here the Zeeman paramagnetic term $\pm\mu_BH$, which requires
summation over both spin orientations in Eq. (\ref{d}), is included.
In the gauge $A_y=Hx$, the solutions are $f_m(x)=\tilde{f}_m
\exp(-\pi Hx^2/\phi_0)$, and $\Delta_m(x)=\tilde{\Delta}_m\exp(-\pi
Hx^2/\phi_0)$, where $\tilde{f}_m$ is expressed via
$\tilde{\Delta}_m$ from Eqs. (\ref{uzl1}) and (\ref{uzl2}). The
solvability condition (\ref{d}) of two linear equations for
$\tilde{\Delta}_1$ and $\tilde{\Delta}_2$ gives the equation for
$H_{c2}$ \cite{ag}, which accounts for interband and intraband
scattering and paramagnetic effects:
    \begin{eqnarray}
    (\lambda_0+\lambda_i)(\ln t+U_+)+ (\lambda_0-\lambda_i)(\ln t +U_-)
    \nonumber \\
     +2w(\ln t+U_+)(\ln t + U_-)=0,
    \label{final}
    \end{eqnarray}
where $t=T/T_{c0}$, and
    \begin{eqnarray}
    \lambda_i=[(\omega_- +\gamma_- )\lambda_-
    -2\lambda_{12}\gamma_{21}-2\lambda_{21}\gamma_{12}]/\Omega_0,
    \label{lambdai} \\
    2\Omega_\pm =\omega_++\gamma_+\pm\Omega_0,\qquad\qquad
    \label{par1} \\
    \Omega_0=[(\omega_{-}+\gamma_{-})^2+4\gamma_{12}\gamma_{21}]^{1/2}, \qquad
    \label{par2} \\
    \omega_\pm=(D_1\pm D_2)\pi H/\phi_0,\qquad\quad
    \label{par3} \\
    U_\pm=\mbox{Re} \psi\left(\frac{1}{2}+\frac{\Omega_\pm +i\mu_BH}{2\pi T}\right)-\psi\left(\frac{1}{2}\right).
    \label{tU}
    \end{eqnarray}
If interband scattering and paramagnetic effects are negligible,
Eqs. (\ref{final})-(\ref{tU}) reduce to a simpler equation
\cite{ag,gk}, which can be presented in the parametric form:
    \begin{eqnarray}
    \ln t=-[U(h)+U(\eta h)+\lambda_0/w]/2+
    \\ \nonumber
    [(U(h)-U(\eta
    h)-\lambda_-/w)^2/4+\lambda_{12}\lambda_{21}/w^2]^{1/2},
    \label{hc2} \\
    H_{c2}=2\phi_0T_cth/D_1,\qquad\qquad
    \label{hh2}
    \end{eqnarray}
where $\eta=D_2/D_1$, and the parameter h runs from $0$ to $\infty$
as T varies from $T_c$ to 0. For equal diffusivities, $\eta=1$, Eq.
(\ref{hc2}) simplifies to the one-gap de-Gennes-Maki equation $\ln
t+U(h)=0$ \cite{dege,whh,maki}.

Now we consider some limiting cases, which illustrate how $H_{c2}$
depends on different parameters. Fig. 5 shows the evolution of
$H_{c2}$ as $g$ increases for fixed $D_1$ and $D_2$ and negligible
paramagnetic effects. Interband scattering reduces the upward
curvature of $H_{c2}(T)$, $H_{c2}(0)$, and $T_c$, while increasing
the slope $H_{c2}^\prime$ at $T_c$. Notice that the significant
changes in the shape of $H_{c2}(T)$ in Fig. 5 occur for weak
interband scattering ($g\ll 1$), which also provides a finite
$H_{c2}(0)$ even if $D_2\to 0$. For example, the high-field films in
Fig. 1 and 2 have $g\simeq 0.045$ and $0.065$, respectively. For
$g\ll 1$, Eq. (\ref{final}) yields the GL linear temperature
dependence near $T_c$:
    \begin{equation}
    H_{c2}=\frac{8\phi_0(T_c-T)}{\pi^2(s_1D_1+s_2D_2)}
    \label{hc2gl}
    \end{equation}
where $T_c$ is given by Eq. (\ref{tcsmall}),
$s_1=1+\lambda_-/\lambda_0$ and $s_2=1-\lambda_-/\lambda_0$. Eq.
(\ref{hc2gl}) is written in the linear accuracy in $g\ll 1$. Higher
order terms in g not only shift $T_c$ but also increase the slope
$H_{c2}^\prime$ at $T_c$, as evident from Fig. 5. For $s_1\sim s_2$,
the slope $H_{c2}^\prime$ is mostly determined by the cleanest band
with the maximum diffusivity. However, because of weak interband
coupling in MgB$_2$, the values of $s_1$ and $s_2$ are very
different. For $\lambda_{11}= 0.81$, $\lambda_{22}= 0.285$,
$\lambda_{12}= 0.119$, $\lambda_{21}= 0.09$ \cite{goluba}, we get
$\lambda_-=\lambda_{11}-\lambda_{22}=0.525$,
$\lambda_0=(\lambda_-^2+4\lambda_{12}\lambda_{21})^{1/2}=0.564$,
thus $s_1=1+\lambda_-/\lambda_0=1.93$,
$s_2=1-\lambda_-/\lambda_0=0.07$. Thus, $H_{c2}^\prime$ is mostly
determined by $D_1$ of the $\sigma$ band. Yet, if the $\sigma$ band
is so dirty that $D_1/D_2 < s_2/s_1\simeq 0.04$, the slope
$H_{c2}^\prime$ is determined by the much cleaner $\pi$ band.

\begin{figure}          
\epsfxsize= 0.85\hsize \centerline{ \vbox{ \epsffile{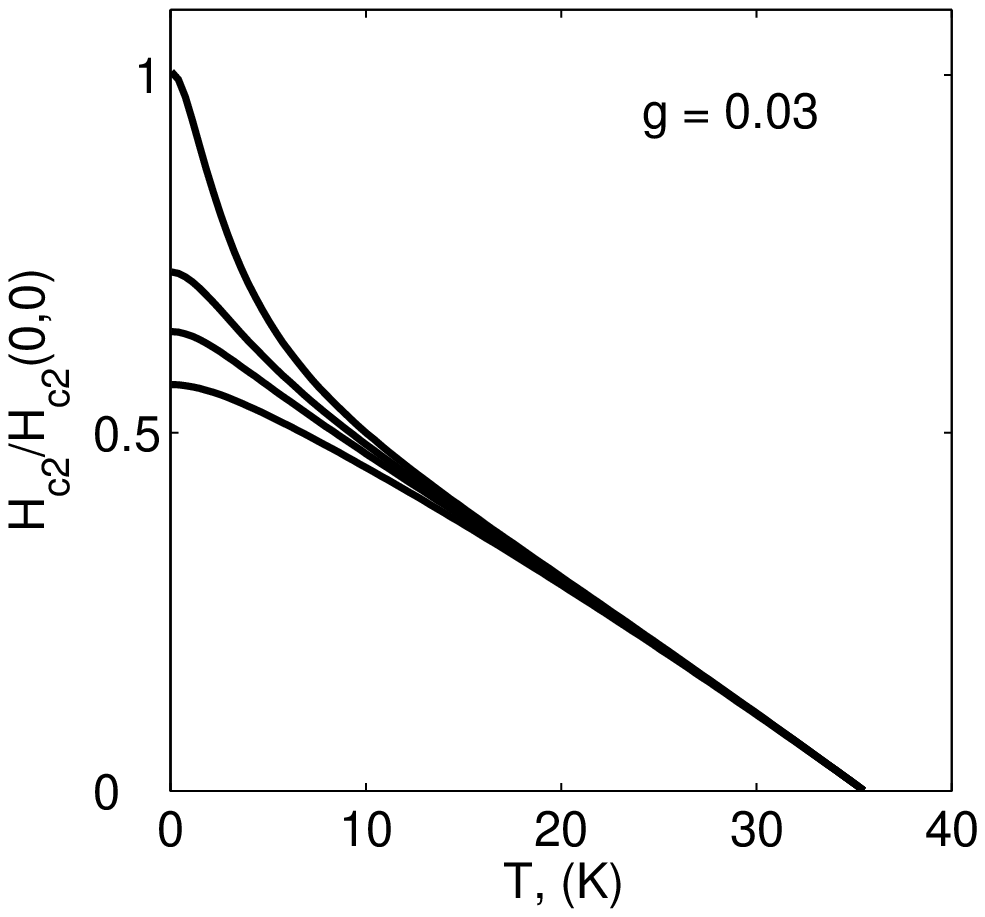} }}
\epsfxsize= 0.85\hsize \centerline{ \vbox{ \epsffile{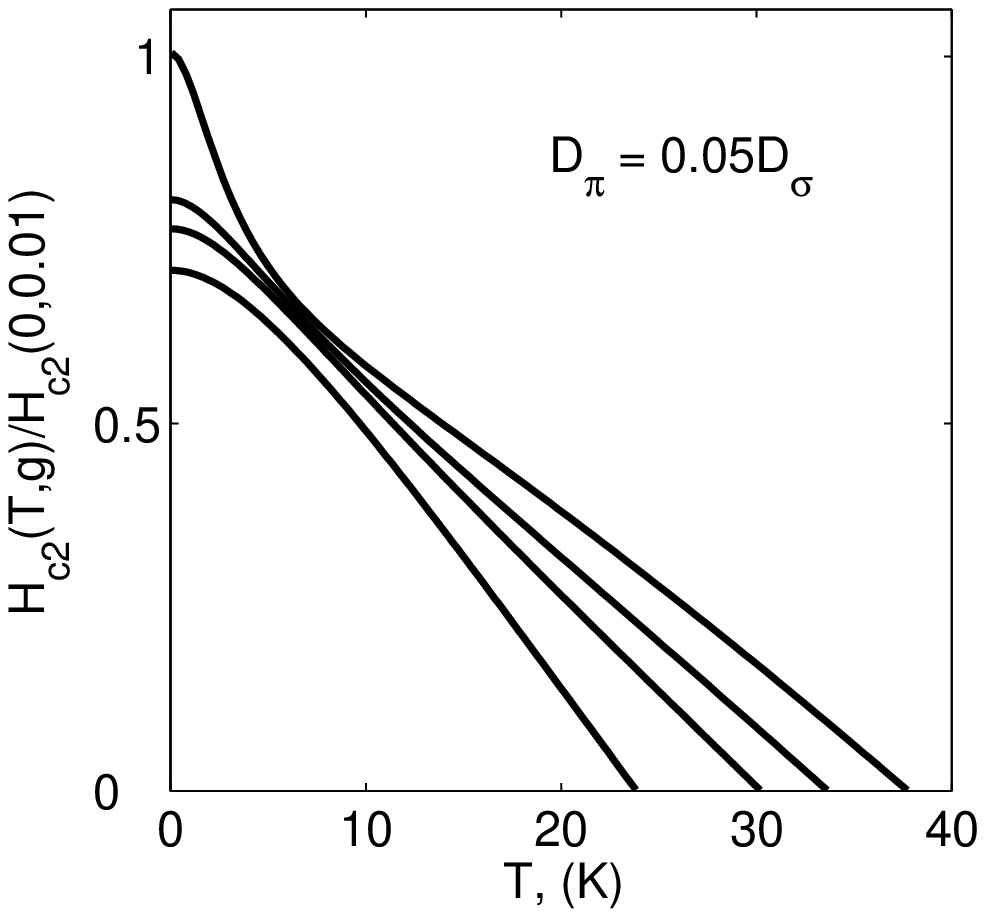} }}
\vskip \baselineskip \caption{Effect of interband scattering and the
diffusivity ratio on the evolution of $H_{c2}(T)$. Upper panel shows
$H_{c2}(T,\eta)$ for the fixed $g=0.03$ and different
$\eta=D_2/D_1$: $0; 0.05; 0.1; 0.5$ (from top to bottom curves).
Lower panel shows $H_{c2}(T,g)$ for the fixed $D_2/D_1=0.05$ and
different $g=0.01; 0.05; 0.1; 0.5$ (from top to bottom curves).
 }
\label{fig.5}
\end{figure}

At low T both the Zeeman and interband scattering terms in Eq.
(\ref{final}) can be essential. Eq. (\ref{final}) reduces to the
following equation for $H_{c2}(0)$:
    \begin{eqnarray}
    (\lambda_0+\lambda_i)\ln\frac{\mu_B^2H_p^2}{\mu_B^2H^2+\Omega_+^2}+
    (\lambda_0-\lambda_i)\ln\frac{\mu_B^2H_p^2}{\mu_B^2H^2+\Omega_-^2} \nonumber \\
    =w\ln\frac{\mu_B^2H_p^2}{\mu_B^2H^2+\Omega_+^2}\ln\frac{\mu_B^2H_p^2}{\mu_B^2H^2+\Omega_-^2}\qquad\qquad
    \label{ok}
    \end{eqnarray}
where $\mu_BH_p = \pi T_{c0}/2\gamma$ is the field of paramagnetic
instability of the superconducting state, and $\ln\gamma = 0.577$.
We first consider the limit $g\to 0$, which defines the maximum
$H_{c2}(0)$ achievable in a dirty two-gap superconductor with no
$T_c$ suppresion. In this case $\Omega_+=\pi D_1H/\phi_0$ and
$\Omega_-=\pi D_2H/\phi_0$, so for $T\ll T_c$, paramagnetic effects
just renormalize intraband diffusivities in Eq. (\ref{ok}):
    \begin{equation}
    D_m\to \tilde{D}_m=\sqrt{D_m^2+D_0^2},
    \label{difren}
    \end{equation}
where $D_0=\mu_B\phi_0/\pi$ is the quantum diffusivity
    \begin{equation}
    D_0=\hbar/2m,
    \label{qdif}
    \end{equation}
and $m$ is the bare electron mass. Eq. (\ref{qdif}) follows from the
basic diffusion relation $l^2=D_0t$, and the energy uncertainty
principle $\hbar^2/2ml^2=\hbar/t$ for a particle confined in a
region of length $l$. For $g=0$, Eq. (\ref{ok}) yields
    \begin{eqnarray}
    H_{c2}(0)=\frac{\phi_0T_c}{2\gamma\sqrt{ \tilde{D}_1 \tilde{D}_2}}\exp(\frac{f}{2}),\qquad\qquad
    \label{hc2o} \\
    f=\left( \frac{\lambda_0^2}{w^2}+\ln^2\frac{\tilde{D}_2}{\tilde{D}_1}+
    \frac{2\lambda_-}{w}\ln\frac{\tilde{D}_2}{\tilde{D}_1}\right)^{1/2}-\frac{\lambda_0}{w}.\quad
    \label{foo}
    \end{eqnarray}
If $D_0\ll D_m$, Eqs. (\ref{hc2o})-(\ref{foo}) reduce to the result
of Ref. \cite{ag}, and for the symmetric case,
$\tilde{D}_1=\tilde{D}_2$, Eqs. (\ref{hc2o})-(\ref{foo}) give the
one-band result $H_{c2}(0)=\phi_0T_c/2\gamma \tilde{D}$ \cite{whh}.
However for $\tilde{D}_1\neq \tilde{D}_2$, $H_{c2}(0)$ can be much
higher than $H_{c2}(0)=0.69H_{c2}^\prime T_c$. Indeed, if the
effective diffusivities, $\tilde{D}_1$ and $\tilde{D}_2$ are very
different, Eqs. (\ref{hc2o})-(\ref{foo}) yield
    \begin{eqnarray}
    H_{c2}(0)=\frac{\phi_0T_c}{2\gamma \tilde{D}_2}e^{-(\lambda_- +\lambda_0)/2w},
    \qquad \tilde{D}_2\ll \tilde{D}_1e^{-\frac{\lambda_0}{w}},
    \label{o1} \\
    H_{c2}(0)=\frac{\phi_0T_c}{2\gamma \tilde{D}_1}e^{-(\lambda_0 -\lambda_-)/2w},
    \qquad \tilde{D}_1\ll \tilde{D}_2e^{-\frac{\lambda_0}{w}}.
    \label{o2}
    \end{eqnarray}
Thus, $H_{c2}(0)$ is determined by the {\it minimum} effective
diffusivity, but unlike the limit $D_0\to 0$, $H_{c2}(0)$ remains
finite even for $D_1\to 0$ or $D_2\to 0$. In fact, if both $D_1\ll
D_0$ and $D_2\ll D_0$, we return to the symmetric case
$\tilde{D}_1=\tilde{D}_2$, for which Eqs. (\ref{hc2o})-(\ref{foo})
yield the result of the one-gap dirty limit theory \cite{maki}
    \begin{equation}
    H_{c2}(0)\to H_p=\phi_0T_c/2\gamma D_0 =\pi T_c/2\gamma\mu_B
    \label{hc2max}
    \end{equation}
For a one-band superconductor, Eq. (\ref{hc2max}) can also be
written as the paramagnetic pairbreaking condition,
$\mu_BH_p=\Delta(0)/2$, where $\Delta(0)=\pi T_c/\gamma$ is the
zero-temperature gap. For two-band superconductors, the meaning of
$H_p$ is less transparent, yet the maximum $H_p$ expressed via $T_c$
is given by the same Eq. (\ref{hc2max}) as for one-band
superconductors.

\begin{figure}          
\epsfxsize= 0.85\hsize \centerline{ \vbox{ \epsffile{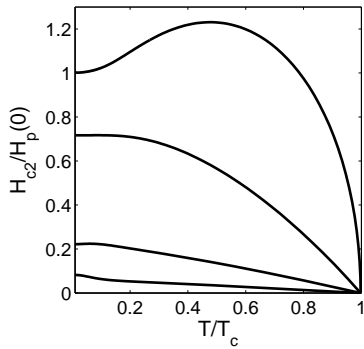} }}
\vskip \baselineskip \caption{Crossover from the orbitally to
paramagnetically limited $H_{c2}(T)$ calculated from Eqs.
(\ref{hc2}) and (\ref{up1})-(\ref{parr}) for $D_2=0.05D_1$ and
$D_1/D_0 = 0, 1, 5, 20$ from top to bottom curves, respectively.
 }
\label{fig.6}
\end{figure}

Finally we consider how paramagnetic effects affect the shape of
$H_{c2}(T)$ in the limit $g\to 0$. This case is described by Eq.
(\ref{hh2}) modified as follows:
    \begin{eqnarray}
    U(h)\to \mbox{Re}\psi[1/2+h(i+p)] - \psi(1/2)\\ \label{up1}
    U(\eta h)\to\mbox{Re}\psi[1/2+h(p\eta+i)] - \psi(1/2)\\
    \label{up2}
    H_{c2}=H_{c2}=2\phi_0T_cth/D_0,\qquad\qquad
    \label{parr}
    \end{eqnarray}
where $p=D_1/D_0$, and $\eta=D_2/D_1$. Fig. 6 shows how $H_{c2}(T)$
evolves from the orbitally-limited $H_{c2}(T)$ with an upward
curvature at $D_1\gg D_2$ to the paramagnetically-limited
$H_{c2}(T)$ of a one-gap superconductor for $D_1<D_0$ \cite{par3}.
The nonmonotonic dependence of $H_{c2}(T)$ in Fig. 6 indicates the
first order phase transition, similar to that in one-gap
superconductors.

\section{Thin films in a parallel field}

$H_{c2}$ can be significantly enhanced in thin films or multilayers,
in which MgB$_2$ layers are separated by nonsuperconducting layers.
It is well known that in a thin film of thickness $d<\xi$ in a
parallel field, $H_{c2}^{(f)}=2\sqrt{3}H_{c2}\xi/d$ can be higher
than the bulk $H_{c2}=\phi_0/2\pi\xi^2$ \cite{layer2,tink}. Let us
see how this result is generalized to two-gap superconductors. For a
thin film of thickness $d<\mbox{max}(\xi_1, \xi_2)$, the functions
$f_1$ and $f_2$ are nearly constant, so integrating Eqs.
(\ref{uzl1}) and (\ref{uzl2}) over x with $\partial_xf(\pm d/2)=0$,
results in two linear equations for $f_1$ and $f_2$ with $\Pi^2 =
(\pi Hd/\phi_0)^2/3$. Thus, we obtain the previous Eq.
(\ref{final})-(\ref{tU}) in which one should make the replacement
    \begin{equation}
    \omega_\pm^{(f)}\to (\pi Hd/\phi_0)^2(D_1\pm D_2)/6
    \label{film}
    \end{equation}
We first consider the case of negligible interband scattering and
paramagnetic effects. Then Eq. (\ref{final}) and (\ref{film}) give
the square-root temperature dependence near $T_c$
    \begin{equation}
    H_{c2}^{(f)}=\frac{4\phi_0\sqrt{3T_c(T_c-T)}}{\pi^{3/2}d(s_1D_1+s_2D_2)^{1/2}}
    \label{filmTc}
    \end{equation}
characteristic of thin films \cite{tink} instead of the bulk GL
linear dependence (\ref{hc2gl}). From Eqs. (\ref{hc2o}) and
(\ref{film}) we can also obtain $H_{c2}^{(f)}(0)$ for $D_0\ll D_m$:
    \begin{equation}
    H_{c2}^{(f)}(0)=\frac{\phi_0}{d}\left(\frac{3T_c}{\pi\gamma}\right)^{1/2}\frac{\exp(f/4)}{(D_1D_2)^{1/4}}
    \label{filmo}
    \end{equation}
Next we consider the crossover to the paramagnetic limit in thin
films at low temperatures. For neglect interband scattering, the
expressions $\mu_B^2H^2+\Omega^2 $ under the logarithms in Eq.
(\ref{ok}) become $\mu_B^2H^2+(\pi Hd/\phi_0)^4D^2/36$. Substituting
here $H_{c2}^{(f)}\sim \phi_0/\xi d$, we conclude that paramagnetic
effects become essential if
    \begin{equation}
    \mbox{min}(D_1, D_2) < D_0\xi/d.
    \label{Dfilm}
    \end{equation}
Thus, reducing the film thickness extends the region of the
parameters where $H_{c2}$ is limited by the paramagnetic effects
rather than by impurity scattering.

\section{Anisotropy of $H_{c1}$ and $H_{c2}$}

For anisotropic one-gap superconductors, the angular dependence of
the lower and the upper critical fields is given by \cite{gork,bul}
    \begin{equation}
    H_{c1}(\alpha,T)=\frac{H_{c1}(0,T)}{R(\alpha)}, \quad
    H_{c2}(\alpha,T)=\frac{H_{c2}(0,T)}{R(\alpha)}
    \label{onegapanis}
    \end{equation}
where $R(\alpha)=(\cos^2\alpha+\epsilon\sin^2\alpha)^{1/2}$,
$\epsilon = m_{ab}/m_c$. Here the anisotropy parameter
$\Gamma(T)=H_{c2}^{||}/H_{c2}^{\perp}={\epsilon}^{-1/2}$ is
independent of T for both $H_{c1}$ and $H_{c2}$. By contrast,
$\Gamma_2(T)=H_{c2}^{||}/H_{c2}^{\perp}$ for MgB$_2$ single crystals
increases from $\sim 2-3$ at $T_c$ to $5-6$ at $T\ll T_c$, but
$\Gamma_1(T)=H_{c1}^{||}/H_{c1}^{\perp}$ decreases from $\simeq 2-3$
to $\simeq 1$ as T decreases
\cite{a1,a2,a3,a4,a5,a6,a7,a8,a9,a10,a11}. This behavior was
explained by the two-gap theory in the clean limit
\cite{vgk,dahm,lanis1,lanis2}.

The dirty limit is more intricate in the sense that $\Gamma(T)$ can
either increase or decrease with T, depending on the diffusivity
ratio $D_2/D_1$. However, the physics of this dependence is rather
transparent and can be understood using the bilayer toy model as
discussed in the Introduction. Indeed, for very different $D_1$ and
$D_2$, both the angular and the temperature dependencies of
$H_{c2}(\alpha,T)$ are controlled by cleaner band at high T and by
dirtier band at lower T. For instance, if $D_2\ll D_1$, the high-T
part of $H_{c2}(\alpha,T)$ is determined by the anisotropic $\sigma$
band, while the low-T part is determined by the isotropic $\pi$
band. In this case $\Gamma(T)$ decreases as T decreases, as
characteristic of dirty MgB$_2$ films represented in Figs. 1 and 2.
If the $\pi$ band is cleaner than the $\sigma$ band, $\Gamma(T)$
increases as T decreases, similar to single crystals.

For the field ${\bf H}$ inclined with respect to the c-axis, the
first Landau level eigenfunction no longer satisfies Eqs.
(\ref{uzl1}), (\ref{uzl2}) and (\ref{d}). In this case $f_m(\omega,
{\bf r})$ are to be expanded in full sets of eigenfunctions for all
Landau levels, and $H_{c2}$ becomes a root of a matrix equation
$\hat{M}(H_{c2})=0$ \cite{ag,gk}. As shown in Ref. \cite{ag}, this
matrix equation for $H_{c2}$ greatly simplifies for the moderate
anisotropy characteristic of dirty MgB$_2$ for which all formulas of
the previous section can also be used for the inclined field as well
by replacing $D_1$ and $D_2$ with the angular-dependent
diffusivities $D_1(\alpha)$ and $D_2(\alpha)$ for {\it both bands}:
    \begin{equation}
    D_m(\alpha)= [D_m^{(a) 2}\cos^2\alpha+D_m^{(a)}D_m^{(c)}\sin^2\alpha]^{1/2}
    \label{anis}
    \end{equation}
In terms of the bilayer model shown in Fig. 1, Eq. (\ref{anis}) just
means that Eq. (\ref{onegapanis}) should be applied separately for
each of the films. For $g=0$, Eqs. (\ref{hc2gl}) and (\ref{anis})
determine the angular dependence of $H_{c2}(\alpha)$ near $T_c$, and
the London penetration depth $\Lambda_{\alpha\beta}$ is given by
\cite{ag}
    \begin{equation}
    \Lambda^{-2}_{\alpha\beta}=\frac{4\pi^4}{\phi_0^2}\bigl[N_1D_1^{\alpha\beta}\Delta_1\tanh\frac{\Delta_1}{2T}+
    N_2D_2^{\alpha\beta}\Delta_2\tanh\frac{\Delta_2}{2T}\bigr]
    \label{lond}
    \end{equation}
Eqs. (\ref{hc2gl}), (\ref{anis}), and (\ref{lond}) show that the
one-gap scaling (\ref{onegapanis}) breaks down because the behavior
of $H_{c1}(\alpha,T)$ is mostly controlled by the cleaner band for
all T, while the behavior of $H_{c2}(\alpha,T)$ is determined by the
cleaner band at higher T, and by the dirtier band at lower T. Thus,
$\Gamma_1(T)$ and $\Gamma_2(T)$ for $H_{c1}$ and $H_{c2}$ in the
two-gap dirty limit are different. Temperature dependencies of
$\Gamma(T)$ were calculated in Refs. \cite{ag,gk}.

Eqs. (\ref{anis}) and (\ref{final}) describe well both the
temperature and the angular dependencies of $H_{c2}(\alpha,T)$ in
dirty MgB$_2$ films \cite{h6,h9,h10,h11}. Eq. (\ref{anis}) is valid
if the $\sigma$ band is not too anisotropic, and the off-diagonal
elements $M_{mn}\sim \zeta^{m+n}$ are negligible provided that
$\zeta\ll 1$ \cite{ag}. Here
    \begin{equation}
    \zeta=\frac{(\epsilon_1-\epsilon_2)^2\sin^4\alpha}
    {[\sqrt{\cos^2\alpha+\epsilon_1\sin^2\alpha}+\sqrt{\cos^2\alpha+\epsilon_2\sin^2\alpha}]^4},
    \label{anisappl}
    \end{equation}
$\epsilon_1=D_1^{(c)}/D_1^{(ab)}$ and
$\epsilon_2=D_2^{(c)}/D_2^{(ab)}$. For $\epsilon_2=1$, the parameter
$\zeta(\alpha) <0.45$ for a rather strong anisotropy $\epsilon_1 <
0.04$ and $\alpha=\pi/2$. For a stronger anisotropy, the condition
$\zeta(\alpha)\ll 1$ can still hold in a wide range of $\alpha$,
except a vicinity of $\alpha\approx \pi/2$. In this case the
calculation of $H_{c2}(\alpha,T)$ requires a numerical solution of
the matrix equation for $H_{c2}$ \cite{gk}. However, the Usadel
theory can only be applied to dirty MgB$_2$ samples which, contrary
to the assumption of Ref. \cite{gk}, usually exhibit much weaker
anisotropy $(\Gamma_2\simeq 1-2)$ than single crystals. Perhaps,
strong impurity scattering and admixture of interband scattering
reduce the anisotropy of $D_1^{(c)}/D_1^{(ab)}\simeq 0.2-0.3$ as
compared to that of the Fermi velocities $\langle
v^2_c\rangle_\sigma/\langle v^2_{ab}\rangle_\sigma\sim 0.02$
predicted by {\it ab-initio} calculations for single crystals
\cite{anz1}. The moderate anisotropy of $D_1$ in dirty MgB$_2$ makes
the scaling rule (\ref{anis}) a very good approximation, as was
recently confirmed experimentally \cite{h11}.

\section{Ginzburg-Landau equations}

The two-gap GL equations were obtained both for the dirty limit
without interband scattering \cite{ag,golkosh}, and for the clean
limit \cite{zhit}. Here we consider the GL dirty limit, focusing on
new effects brought by interband scattering. For $\gamma_{mm'}=0$,
the Usadel equations near $T_c$ yield
    \begin{equation}
    f_m=\Psi_m/\omega+D_{m\alpha}\Pi^2_\alpha\Psi/2\omega^2-\Psi_m|\Psi_m|^2/2\omega^3,
    \label{fgl}
    \end{equation}
where the principal axis of $D_{\alpha\beta}$ are taken along the
crystalline axis. For weak interband scattering, the free energy
$F=F_0+F_i$ contains the free energy $F_0\{\Psi_1,\Psi_2\}$ for
$\gamma_{mm'}=0$ and the correction $F_i\{\Psi_1,\Psi_2\}$ linear in
$\gamma_{mm'}$. Here $F_0$ does not have first order corrections in
$\gamma_{mm'}$ if $\Psi_m$ satisfies the GL equations, so $F_i$ can
be calculated by substituting Eq. (\ref{fgl}) into Eq. (\ref{Fi})
and expanding $g_m\approx 1-|f_m|^2/2-|f_m|^4/8$:
    \begin{equation}
    F_i=\pi qT\sum_{\omega
    >0}[|f_1-f_2|^2+(|f_1|^2-|f_2|^2)^2/4],
    \label{figl}
    \end{equation}
where $q=(N_1\gamma_{12}+N_2\gamma_{21})/2$. Combining $F_i$ with
$F_0$ in the dirty limit for $g=0$ \cite{ag}, we arrive at the GL
free energy $\int FdV$ for $g\ll 1$:
    \begin{eqnarray}
    F=a_1|\Psi_1|^2+c_{1\alpha}|\Pi_\alpha\Psi_1|^2+b_1|\Psi_1|^4/2
    \nonumber \\
    +a_2|\Psi_2|^2+c_{2\alpha}|\Pi_\alpha\Psi_2|^2+b_2|\Psi_2|^4/2
    \nonumber \\
    -a_i{\mbox Re}(\Psi_1\Psi_2^*)+c_{i\alpha}{\mbox Re}(\Pi_\alpha\Psi_1\Pi_\alpha^*\Psi_2^*)
    \nonumber \\
    -b_i|\Psi_1|^2|\Psi_2|^2+2b_i(|\Psi_1|^2+|\Psi_2|^2){\mbox
    Re}(\Psi_1\Psi_2)
    \label{glf}
    \end{eqnarray}
Here the GL expansion coefficients are given by
    \begin{eqnarray}
    a_1=\frac{N_1}{2}\bigl[\ln\frac{T}{T_1}+\frac{\pi\gamma_{12}}{4T}\bigr],
    \label{a1}\\
    a_2=\frac{N_2}{2}\bigl[\ln\frac{T}{T_2}+\frac{\pi\gamma_{21}}{4T}\bigr],
    \label{a2} \\
    c_{1\alpha}=N_1D_{1\alpha}\bigl[\frac{\pi}{16T}-\frac{7\zeta(3)\gamma_{12}}{8\pi^2T^2}\bigr],
    \label{c1} \\
    c_{2\alpha}=N_2D_{2\alpha}\bigl[\frac{\pi}{16T}-\frac{7\zeta(3)\gamma_{21}}{8\pi^2T^2}\bigr],
    \label{c2}
    \end{eqnarray}
    \begin{eqnarray}
    b_1=N_1\bigl[\frac{7\zeta(3)}{16\pi^2T^2}-\frac{3\pi\gamma_{12}}{384T^3}\bigr],
    \label{b1} \\
    b_2=N_2\bigl[\frac{7\zeta(3)}{16\pi^2T^2}-\frac{3\pi\gamma_{21}}{384T^3}\bigr],
    \label{b2} \\
    a_i=\frac{N_1}{2}\bigl[\frac{\lambda_{12}}{w}+\frac{\pi\gamma_{12}}{4T}\bigr]+
    \frac{N_2}{2}\bigl[\frac{\lambda_{21}}{w}+\frac{\pi\gamma_{21}}{4T}\bigr],
    \label{ai} \\
    c_i=\frac{7\zeta(3)}{(4\pi
    T)^2}(D_1+D_2)(\gamma_{12}N_1+\gamma_{21}N_2),
    \label{ci} \\
    b_i=\frac{\pi}{384T^3}(\gamma_{12}N_1+\gamma_{21}N_2),
    \label{bi}
    \end{eqnarray}
where $T_1=T_{c0}\exp[-(\lambda_0-\lambda_-)/2w]$, and
$T_2=T_{c0}\exp[-(\lambda_0+\lambda_-)/2w]$. The GL equations are
obtained by varying $\int FdV$. I would like to point out the
misprints with wrong signs of $a_i$, $c_1$ and $c_2$ in Eqs. (13),
(14) and (20) in Ref. \cite{ag} (see also Ref. \cite{zhit}).

The first two lines in Eq. (\ref{glf}) are the GL intraband free
energies and the term $a_i{\mbox Re}(\Psi_1\Psi_2^*)$ describes the
Josephson coupling of $\Psi_1$ and $\Psi_2$. Interband scattering
increases $a_1$ and $a_2$, and the interband coupling constant
$a_i$. The net result is the reduction of $T_c$ determined by the
equation $4a_1(T_c)a_2(T_c)=a_i^2$, which reproduces Eq.
(\ref{tcsmall}). Besides the renormalization of $a_m$, $b_m$ and
$c_m$, interband scattering produces new terms, which describe the
mixed gradient coupling and the nonlinear quatric interaction of
$\Psi_1$ and $\Psi_2$. Similar terms were introduced in the GL
theories of heavy fermions \cite{heavyferm} and borocarbides
\cite{sigrist}, and phenomenological models of $H_{c2}$ in MgB$_2$
\cite{asker}. These terms result from interband scattering, so both
$c_i$ and $b_i$ vanish in the clean limit \cite{zhit}. The mixed
gradient terms in Eq. (\ref{glf}) produce interference terms in the
current density ${\bf J} = -c\delta F/\delta {\bf A}$:
    \begin{eqnarray}
    {\bf J}=- [(2c_1\Delta_1^2+c_i\Delta_1\Delta_2\cos\theta){\bf
    Q}_1+ \nonumber \\
    (2c_2\Delta_2^2+c_i\Delta_1\Delta_2\cos\theta){\bf Q}_2
    +\nonumber \\
    c_i(\Delta_2\nabla\Delta_1-\Delta_1\nabla\Delta_2)\sin\theta ]2\pi
    c/\phi_0
    \label{jmix}
    \end{eqnarray}
where ${\bf Q}_m=\nabla\theta_m+2\pi{\bf A}/\phi_0$, and
$\theta=\theta_1-\theta_2$. Here ${\bf J}$ is no longer the sum of
independent contributions of two bands, because phase gradients in
one band produce currents in the other. Moreover, ${\bf J}$ acquires
new $\cos\theta$ terms and the peculiar $\sin\theta$ interband
Josephson-like contribution for inhomogeneous gaps. For currents
well below the depairing limit, both bands are phase-locked
($\theta=0$), and Eq. (\ref{jmix}) defines the London penetration
depth $\Lambda^2=c\phi_0Q/8\pi^2|J|$:
    \begin{equation}
    \Lambda=\phi_0/4\pi[2\pi(c_1\Delta_1^2+c_i\Delta_1\Delta_2+c_2\Delta_2^2)]^{1/2}
    \label{lond}
    \end{equation}
where $c_1$, $c_2$ and $c_i$ depend on the field orientation
according to Eq. (\ref{anis}). Eq. (\ref{fgl}) can be used to
calculate $H_{c2}^\perp(T)$ from the linearized GL equations, which
give $H_{c2}$ as a solution of the quadratic equation \cite{sigrist}
    \begin{equation}
    4\left[\frac{2\pi c_1H}{\phi_0}+a_1\right]\left[\frac{2\pi
    c_2H}{\phi_0}+a_2\right]=\left[a_i+\frac{2\pi
    c_iH}{\phi_0}\right]^2
    \label{glhh}
    \end{equation}
which reduces to Eq. (\ref{hc2gl}) near $T_c$ to the linear accuracy
in $\gamma_{mm'}$. However, GL calculations of $H_{c2}(T)$ in
MgB$_2$ beyond the linear $T_c-T$ term \cite{asker,zhit1} have a
rather limited applicability, since $a_1(T)$ and $a_2(T)$ change
signs at very different temperatures $T_1$ and $T_2$. For
$\lambda_{mn}$ of Ref. \cite{goluba}, $T_1\sim 0.9T_{c0}$ and
$T_2\sim 0.1T_{c0}$ so higher order gradient terms (automatically
taken into account in the Eliashberg/Eilenberger/Usadel based
theories) become important. For example, at $T\approx T_1$ where
$a_2(T_1)\gg a_1(T)$, retaining the first gradient term $\propto
c_2$ requires taking into account a next order term $\sim H^2$ in
the first brackets in Eq. (\ref{glhh}), which is beyond the GL
accuracy. Thus, applying the GL theory in a wider temperature range
\cite{asker,zhit1} makes it a procedure of unclear accuracy, which
can result in a spurious upward curvature in $H_{c2}(T)$ not always
present in a more consistent theory (for example, in the dirty limit
at $D_1\simeq D_2$). In addition, the anisotropy of $D_1(\alpha)$
may further limit the applicability of the GL theory for ${\bf
H}||ab$, as for $c_1\gg c_2$ higher order gradient terms in the
$\pi$ band become important \cite{golkosh}.

\section{Discussion}

The remarkable ten-fold increase of $H_{c2}(T)$ in C-doped MgB$_2$
films \cite{h6,h7,h8,h9,h10} has brought to focus new and largely
unexplored physics and materials science of two-gap superconducting
alloys. Moreover, the observations of $H_{c2}$ close to the BCS
paramagnetic limit poses the important question of how far can
$H_{c2}$ be further increased by alloying. This possibility may be
naturally built in the band structure of MgB$_2$, which provides
weak interband coupling and weak interband scattering, thus allowing
MgB$_2$ to be alloyed without strong suppression of $T_c$. For
example, for the C-doped MgB$_2$ film shown in Fig. 1, $\rho_n$ was
increased from $\simeq 0.4\mu\Omega$cm to $560\mu\Omega$cm, yet
$T_c$ was only reduced down to $35$K \cite{h9}. It is the weakness
of interband scattering, which apparently makes it possible to take
advantage of very dirty $\pi$ band to significantly boost $H_{c2}$
in carbon-doped films which typically have $D_\pi\sim 0.1D_\sigma$.
The reasons why scattering in the $\pi$ band of C-doped MgB$_2$
films is so much stronger than in the $\sigma$ band has not been
completely understood, but another immediate benefit for high-field
magnet applications \cite{ap1} is that carbon alloying significantly
reduces the anisotropy of $H_{c2}$ down to $\Gamma(T)\simeq 1-2$.

Despite many yet unresolved issues concerning the two-gap
superconductivity in MgB$_2$ alloys, $H_{c2}$ of C-doped MgB$_2$ has
already surpassed $H_{c2}$ of Nb$_3$Sn (see Fig. 1). Given the
intrinsic weakness of interband scattering, which enables tuning
MgB$_2$ by selective atomic substitutions on Mg and B sites, there
appear to be no fundamental reasons why $H_{c2}$ of MgB$_2$ alloys
cannot be pushed further up toward the strong-coupling paramagnetic
limit (\ref{paraeli}). Thus, understanding the mechanisms of intra
and interband impurity scattering in carbon-doped MgB$_2$, and the
competition between scattering and doping effects becomes an
important challenge for the computational physics. For instance, it
remains unclear why the multiphased C-doped HPCVD grown films
\cite{penn} exhibit higher $H_{c2}$ and weaker $T_c$ suppression
\cite{h9} than uniform carbon solid solutions \cite{h7,carb1,carb2}.
This unexpected result may indicate other extrinsic mechanisms of
$H_{c2}$ enhancement, which are not accounted by the simple two-gap
theory presented here. Among those may be effects of electron
localization or strong lattice distortions in multiphased C-doped
films which can manifest themselves in the buckling of the Mg planes
observed in the dirty fiber-textured MgB$_2$ films shown in Fig. 2
\cite{h6}. Such buckling may enhance scattering in the $\pi$ band
formed by out-of-plane $p_z$ boron orbitals.

Recently significant enhancements of vortex pinning and critical
current densities $J_c$ in MgB$_2$
\cite{jc1,jc2,jc3,jc4,jc5,jc6,jc7} has been achieved, particularly
by introducing SiC \cite{jc3} and ZrB$_2$ \cite{jc6} nanoparticles.
Given these promising results combined with weak current blocking by
grain boundaries \cite{dcl}, the lack of electromagnetic granularity
\cite{mo}, and very slow thermally-activated flux creep
\cite{creep1,creep2}, it is not surprising that MgB$_2$ is being
regarded as a strong contender of traditional high-field magnet
materials like NbTi and Nb$_3$Sn. Despite these achievements, a
detailed theory of pinning in MgB$_2$ is trill lacking. Such theory
should take into account a composite structure of the vortex core,
which consists of concentric regions of radius $\xi_\sigma$ and
$\xi_\pi$ where $\Delta_\sigma(r)$ and $\Delta_\pi(r)$ are
suppressed \cite{vort1,vort2,vort3,vort4,vort5,vort6}. For example,
in MgB$_2$ single crystals the larger vortex cores in the $\pi$ band
start overlapping above the "virtual upper critical field"
$H_v=\phi_0/2\pi\xi_\pi^2\sim 0.5T$, causing strong overall
suppression of $\Delta_\pi$ well below $H_{c2}$ \cite{vort2,vort3}.
This effect can reduce $J_c$ at $H>H_v$, however both $H_v$ and
$H_{c2}$ can be greatly increased by appropriate enhancement of
impurity scattering in MgB$_2$ alloys.

Recently there has been an emerging interest in microwave response
of MgB$_2$ \cite{rf1,rf2,rf3} and a possibility of using MgB$_2$ in
resonant cavities for particle accelerators \cite{rf4,rf5}. These
issues require understanding nonlinear electrodynamics and current
pairbreaking in two-gap superconductors \cite{kunchur,nicol}, in
particular, band decoupling and the formation of interband phase
textures at strong rf currents \cite{text1,text2}.

This work was partially supported by in-house research program at
NHMFL. NHMFL is operated under NSF Grant DMR-0084173 with support
from state of Florida.


\begin{references}
\bibitem{note}
This toy model should not be taken too literally, since
$H_{c2}(\alpha)$ in real multilayers in the inclined field can
depend on the film thickness due to the effect of surfaces on the
nucleation of vortices.
\bibitem{agg} $F_i$ of Ref. \cite{text2} should be multiplied by $-1/2$.
\end{references}
\end{document}